\documentclass[a4paper,11pt]{article}
\pdfoutput=1
\usepackage{jheppub}
\usepackage[T1]{fontenc}

\title{\boldmath Entanglement Entropy of $AdS_5 \times S^5$ with massive flavors}
\author[a]{Sen Hu,}
\author[a]{Guozhen Wu}
\affiliation[a]{Wu Wen-Tsun Key Lab of Mathematics of Chinese Academy of Sciences,\\School of Mathematical Sciences,\\University of Science and Technology of China, Hefei, Anhui 230026, China}
\emailAdd{shu@ustc.edu.cn}
\emailAdd{kcwoo@mail.ustc.edu.cn}

\abstract{We consider backreacted $AdS_5 \times S^5$ coupled with $N_f$ massive flavors introduced by D7-branes. The backreacted geometry is in the Veneziano limit with fixed $N_f/N_c$. By dividing one of the directions into a line segment with length $l$, we get two subspaces. Then we calculate the entanglement entropy between them. With the method of~\cite{1}, we are able to find the cut-off independent part of the entanglement entropy and finally find that this geometry shows no phase transition as the case in pure $AdS_5 \times S^5$.}

\begin{document}
\maketitle
\flushbottom

\section{Introduction}
\label{sec:1}

Entanglement entropy measures in a total system how closely a subsystem A would be tangled with another subsystem B. Especially, if an observer locates in A and gets no access to its complement B, the entanglement entropy is the quantity that shows how much information the observer can get from B. In this case, the subsystem B is like a black hole which lies inside the event horizon which is the boundary separates A from B.

The entanglement entropy of a system with subsystem A and its complement B is defined as:
\begin{equation}
S_A=-tr\rho_A \log\rho_A\,,
\end{equation}
where $S_A$ is the entanglement entropy of subsystem A and $\rho_A$ is the reduced density matrix. Usually, the entanglement entropy (EE) is computed with the method of replica trick which makes it a hard work to calculate.
In \cite{2,3}, Ryu and Takayanagi provide a powerful weapon to compute the EE holographically which is known as the area law. The authors show that the EE of the subsystem A is proportional to the area of the minimal surface whose boundary is given by $\partial A$. The area law reads
\begin{equation}
S_A=\frac{Area(\gamma_A)}{4G_N^{d+2}}\,,
\end{equation}
where $\gamma_A$ is the minimal surface which coincides with the boundary of A in the bulk space of the AdS.

Generally, the EE is divergent in the UV region, and the divergent term is dependent of the UV cut-off. A further move is the proposal of \cite{1}, the authors generalize the work of Ryu and Takayanagi and apply the area law to non-conformal field theories. They also develop a method to find the cut-off independent part of the EE by comparing the difference of the EE of the connected minimal surface with the disconnected one. Also in~\cite{1}, the authors find a first-order phase transition phenomenon which can be related to the confinement/deconfinement phase transition.

In this paper, we will consider the model of $SU(N_c)$ $\mathcal{N} =4$ SYM with $N_f$ massive flavors in the Veneziano limit, where $N_f, N_c \rightarrow \infty,N_f/N_c$ is fixed, in zero temperature\footnote{The original study of the related computation can be found in \cite{5} in which the authors studied the three-dimensional ABJM Chern-Simons matter theory with unquenched massive flavors. The background geometry is of the form $AdS_4 \times {\mathcal{M}}_6$. We would like to thank Niko Jokela for pointing this out to us.}. In D3/D7 defect system, D7 branes separate from the D3 branes in the radial direction, which is related to massive flavors \cite{4}. The case we will study is the straight belt with length $l$. Considering a strip with length $l$ which will divide the $d$ dimensional space into two regions, we will calculate the EE between them holographically. After subtracting the UV cut-off dependent part, we can get the EE which is only related to the entanglement length $l$. The result of this paper shows that the entanglement entropy of the disconnected minimal surface is always larger than that of the connected one, thus the connected case is preferred. Like the pure AdS case, there exists no phase transition between the connected phase and the disconnected phase.

The structure of this paper is as follows. In section \ref{sec:2}, we will introduce the work of Ryu and Takayanagi \cite{2,3}, and we will calculate the entanglement length and EE of $AdS_5\times S^5$. Then we will introduce the work of \cite{1}, which will be used to remove the UV cut-off dependent part of the EE, and show that there is no phase transition in pure $AdS_5\times S^5$. In section \ref{sec:3}, we will summarize the work of \cite{6}, the D3/D7 plasma solution with massive flavors, which will be used later as the model we study. For later use, we will change the radial coordinate $\rho$ to $r$ as what the authors of \cite{6} did in the massless flavor case. After changing the radial coordinate, we get the metric of the D3/D7 geometry with backreaction in zero temperature. Since we get all the methods we need, we can now calculate the entanglement length and EE of D3/D7 backreacted system in zero temperature with the shape of a slab in section \ref{sec:4}. Similar results can be found in \cite{7}.  We find that as the EE of disconnected minimal surface is always larger than that of the connected one, the D3/D7 backreacted system shows no phase transition phenomenon and we will compare this result to the pure $AdS_5\times S^5$.
In section \ref{sec:5}, we will make conclusion on the results and show some possible work in the future.

\section{Holographic entanglement entropy}
\label{sec:2}
\subsection{Entanglement entropy of strip in $AdS_{d+2}$}
\label{sec:2.1}
In this subsection we will give a quick review of the work in \cite{3} (see also \cite{2,14}).

Ryu and Takayanagi study two different kinds of subsystem A in $AdS_{d+2}$ space: the strip $A_s$ and the disk $A_d$ by calculating the EE. We will only concentrate on the strip one since it is related to the work of \cite{1}.

The Poincare metric of $AdS_{d+2}$ reads:
\begin{equation}
{ds}^2=R^2z^{-2}({dz}^2-{dx}^{2}_{0}+\sum^{d}_{i=1}dx^{2}_{i})\,.
\end{equation}
We will consider A with the shape of straight belt with length $l$:
\begin{equation*}
A_S=\{x_i\mid x_1\in[-\frac{l}{2},\frac{l}{2}],x_{2,3,\cdots,d}\in[-\infty, \infty]\}\,.
\end{equation*}
The area of an $n$ dimension submanifold $N\subset M$ can be written as:
\begin{equation}
Area(N)=\int_{N}d^nx\sqrt{|g_N|}\,.
\end{equation}
As for our straight belt, the area action is then:
\begin{equation}
Area=R^dL^{d-1}\int^{\frac{l}{2}}_{-\frac{l}{2}}dx\frac{\sqrt{1+(\frac{dz}{dx})^2}}{z^d}\,.
\end{equation}
Let $x_1 =x$, and $z$ is the function of $x$: $z=z(x)$. In order to find the minimal surface of the strip $A_s$, we need to minimize the area action. Regarding $x$ as time, the Hamiltonian which is independent of time reads:
\begin{equation}
\mathcal{H}=z^{\prime}\frac{d\mathcal{L}}{dz^{\prime}}-\mathcal{L}=\frac{(z^{\prime})^2}{z^d\sqrt{(z^{\prime})^2+1}}-\frac{\sqrt{(z^{\prime})^2+1}}{z^d}\,.
\end{equation}
Notice that the Hamiltonian is a constant, setting it as $\mathcal{H}=-\tilde{z}^{-d}$, where $\tilde{z}$ satisfies ${z^{\prime}|}_{z=\tilde{z}}=0$.
Now we get the relation between $z$ and $x$:
\begin{equation}
\frac{dz}{dx}=\frac{\sqrt{\tilde{z}^{2d}-z^{2d}}}{z^d}\,.
\end{equation}
The length $l$ with respect to the turning point $\tilde{z}$ is given by:
\begin{equation}
\frac{l}{2}=\int^{\tilde{z}}_{0}dz\frac{z^d}{\sqrt{\tilde{z}^{2d}-z^{2d}}}=\frac{\sqrt{\pi}\Gamma(\frac{d+1}{2d})}{\Gamma(\frac{1}{2d})}\tilde{z}\,,
\end{equation}
and the area is:
\begin{equation}
Area_{A_S}=\frac{2R^d}{d-1}\left(\frac{L}{a}\right)^{d-1}-2IR^d\left(\frac{L}{\tilde{z}}\right)^{d-1}\,,
\end{equation}
where $a$ is the UV cut-off and $I$ is the constant given by:
\begin{equation}
I=-\frac{\sqrt{\pi}\Gamma(\frac{1-d}{2d})}{2d\Gamma(\frac{1}{2d})}\,.
\end{equation}
The entanglement entropy of the strip is:
\begin{equation}
\label{eq:2.9}
S_{A_S}=\frac{1}{4G^{d+2}_{N}}\left(\frac{2R^d}{d-1}\left(\frac{L}{a}\right)^{d-1}-\frac{2^d\pi^{\frac{d}{2}}R^d}{d-1}\left(\frac{\Gamma(\frac{d+1}{2d})}{\Gamma(\frac{1}{2d})}\right)^d\left(\frac{L}{l}\right)^{d-1}\right)
\end{equation}

Notice that the expression of EE is divergent in the UV region since $a\rightarrow0$, and the divergent part of leading order is independent of the length $l$ but dependent of the UV cut-off $a$. The second term of (\ref{eq:2.9}) is finite and dependent of the length.

\subsection{Calculation of entanglement entropy}
\label{sec:2.2}
In this subsection we will summarize the work of \cite{1} briefly (see also \cite{8}). The authors generalize the work of Ryu and Takayanagi to confining large $N_c$ theories. The generalized area law is given by:
\begin{equation}
S_A=\frac{1}{4G^{(10)}_{N}}\int d^8\sigma e^{-2\Phi}\sqrt{G^{(8)}_{ind}}\,.
\end{equation}
Let us consider the string frame metric of the gravitational background first:
\begin{equation}
{ds}^{2}_{10}=\alpha(r)(\beta(r){dr}^2+{dx}^{\mu}{dx}_{\mu})+g_{ij}{dy}^i{dy}^j\,,
\end{equation}
where $r$ is the radial coordinate ranging from $r_0$ to $\infty$, $x^\mu (\mu=0,1,\cdots,d)$ parameterize~$\mathbb{R}^{d+1}$, $y^i (i=d+2,\cdots, 9)$ parameterize the $8-d$ internal manifold.
The volume of the internal manifold reads:
\begin{equation}
V_{int}=\int\prod^{8-d}_{i=1}dy^i\sqrt{\det g}\,.
\end{equation}
As mentioned above, we will only consider the case of a strip. Define a new function:
\begin{equation}
H(r)=e^{-4\Phi}V^{2}_{int}\alpha^d\,.
\end{equation}

Following the same procedure in section \ref{sec:2.1}, the EE of the straight belt with length $l$ is:
\begin{equation}
\frac{S_A}{V_{d-1}}=\frac{1}{4G^{(10)}_{N}}\int^{\frac{l}{2}}_{-\frac{l}{2}}dx\sqrt{H(r)}\sqrt{1+\beta(r)(\partial_xr)^2}\,.
\end{equation}
The entanglement length is given by:
\begin{equation}
\label{eq:2.15}
l(\tilde{r})=2\sqrt{H(\tilde{r})}\int^{\infty}_{\tilde{r}}\frac{dr\sqrt{\beta(r)}}{\sqrt{H(r)-H(\tilde{r})}}\,,
\end{equation}
where $\tilde{r}$ is the minimal value of $r$ which is related the turning point of the minimal surface.
There exist two possibilities of the entanglement surfaces: one is connected and the other is disconnected. The disconnected one consists of two cigar-like surfaces, while the connected one is a tube linking the two cigar-like surfaces.

The EE of the disconnected minimal surface can be written as:
\begin{equation}
S_D(\tilde{r})=\frac{V_{d-1}}{2G^{(10)}_{N}}\int^{r_\infty}_{r_0}dr\sqrt{\beta(r)H(r)}\,.
\end{equation}
The EE of the connected minimal surface is given by:
\begin{equation}
S_C(\tilde{r})=\frac{V_{d-1}}{2G^{(10)}_{N}}\int^{r_\infty}_{\tilde{r}}dr\frac{\sqrt{\beta(r)}H(r)}{\sqrt{H(r)-H(\tilde{r})}}\,.
\end{equation}
Notice that the EE of the connected minimal surface is dependent of $\tilde{r}$, while the disconnected one is not. As is analyzed above, both of them are UV divergent and dependent of their UV cut-off. The difference between them is finite and UV cut-off independent:
\begin{equation}
\label{eq:2.18}
\frac{2G^{(10)}_{N}}{V_{d-1}}(S_C-S_D)=\int^{\infty}_{\tilde{r}}dr\sqrt{\beta(r)H(r)}\left(\frac{1}{\sqrt{1-\frac{H(\tilde{r})}{H(r)}}}\right)-\int^{\tilde{r}}_{r_0}dr\sqrt{\beta(r)H(r)}\,.
\end{equation}
We will find that the result is dependent of the entanglement length $l$. Notice that we successfully remove the divergent part of the EE, and this is the universal term we want since it is independent of the UV cut-off. We also need to find the smaller EE by comparing the connected and disconnected solutions. The difference between the EE of the connected and disconnected entanglement surface is the quantity we will study in the remaining of this paper.

\subsection{Entanglement entropy of $AdS_5\times S^5$}
\label{sec:2.3}
In this subsection we will calculate the entanglement length and entanglement entropy of pure $AdS_5\times S^5$ which will be used later to be compared with the results of $AdS_5\times S^5$ with massive flavors. Similar results can be found in \cite{15}. This subsection serves as an example of the EE computation procedure in section~\ref{sec:2.2}.

The metric of $AdS_5\times S^5$ reads:
\begin{equation}
{ds}^2=\frac{R^2}{r^2}{dr}^2+\frac{r^2}{R^2}{dx}^2_{1,3}+R^2{d\Omega}^{2}_{5}\,.
\end{equation}
The functions to be used later in computing the entanglement length and EE are:
\begin{equation}
\beta(U)=\frac{R^4}{r^4},\quad H(r)=\left(\frac{8\pi^2}{3}\right)^2R^4r^6\,,
\end{equation}
and $\tilde{r}$ ranges from $0$ to $\infty$.
The entanglement length is given by:
\begin{equation}
\label{eq:2.21}
l(\tilde{r})=2R^2\frac{\sqrt{\pi}\Gamma(\frac{2}{3})}{\Gamma(\frac{1}{6})}\frac{1}{\tilde{r}}\,.
\end{equation}
Notice that the entanglement length is divergent at the origin $\tilde{r}=0$.
The EE of the disconnected minimal surface is:
\begin{equation}
\begin{split}
\frac{S_D(\tilde{r})}{V_{int}}&=\frac{1}{2G^{(10)}_{N}}\int^{\infty}_{0}dr\sqrt{\frac{R^4}{r^4}\left(\frac{8\pi^2}{3}\right)^2R^4r^6}\\
&=\frac{8\pi^2}{3}R^4\frac{1}{2G^{(10)}_{N}}\int^{\infty}_{0}dr\sqrt{r^2}\,,
\end{split}
\end{equation}
which diverges at UV region.
The EE of the connected minimal surface is:
\begin{equation}
\begin{split}
\frac{S_C(\tilde{r})}{V_{int}}&=\frac{1}{2G^{(10)}_{N}}\int^{\infty}_{\tilde{r}}\frac{dr\sqrt{\frac{R^4}{r^4}}\left(\frac{8\pi^2}{3}\right)R^2r^6}{\sqrt{r^6-\tilde{r}^{6}}}\\
&=\frac{8\pi^2}{3}R^4\frac{1}{2G^{(10)}_{N}}\int^{\infty}_{\tilde{r}}dr\frac{r^4}{\sqrt{r^6-\tilde{r}^{6}}}\,,
\end{split}
\end{equation}
which also diverges at UV region. Notice that the results are in agreement with the discussion in section \ref{sec:2.2}. The next step we will take is to compare the difference between them by using~(\ref{eq:2.18}):
\begin{equation}
\label{eq:2.24}
\begin{split}
S(\tilde{r})\equiv S_C-S_D&=V_{int}\frac{8\pi^2}{3}R^4\frac{1}{2G^{(10)}_{N}}\left(\int^{\infty}_{\tilde{r}}dr\frac{r^4}{\sqrt{r^6-\tilde{r}^{6}}}-\int^{\infty}_{0}r{dr}\right)\\
&=V_{int}\frac{8\pi^2}{3}R^4\frac{1}{2G^{(10)}_{N}}\left(\tilde{r}^{2}\left(-\frac{\sqrt{\pi}\Gamma(\frac{2}{3})}{\Gamma(\frac{1}{6})}\right)-0\right)\,.
\end{split}
\end{equation}
To simplify the analysis, we will set the constant $V_{int}\frac{8\pi^2}{3}R^4\frac{1}{2G^{(10)}_{N}}$ equal to 1. We choose to keep the term 0 in (\ref{eq:2.24}) for the reason that after regularization which means that after subtracting the UV divergent part the EE of the disconnected minimal surface is 0. The counterterm used for regularization is $a^2$, where $a\rightarrow+\infty$. Notice that the result of~(\ref{eq:2.24}) is finite since the divergent parts of the connected and disconnected entanglement surface coincide, which is in agreement with the result in section~\ref{sec:2.2}.

One more step we will take is to express $S_C-S_D$ as the function of entanglement length $l$. From (\ref{eq:2.21}), we get that:
\begin{equation}
\label{eq:2.25}
S(l)=-\left(2R^2\frac{\sqrt{\pi}\Gamma(\frac{2}{3})}{\Gamma(\frac{1}{6})}\right)^2\frac{1}{l^2}\,.
\end{equation}
The result of $S(l)$ is always negative. From (\ref{eq:2.25}), we can get the result that since the EE of the disconnected minimal surface is always larger than that of the connected one, hence the EE of connected minimal surface is preferred. There exists no phase transition between them. In addition, the EE of connected entanglement surface is dependent of the entanglement length $l$ while the disconnected one is not.
\begin{figure}[bhtp]
\centering
\includegraphics[width=.45\textwidth,clip]{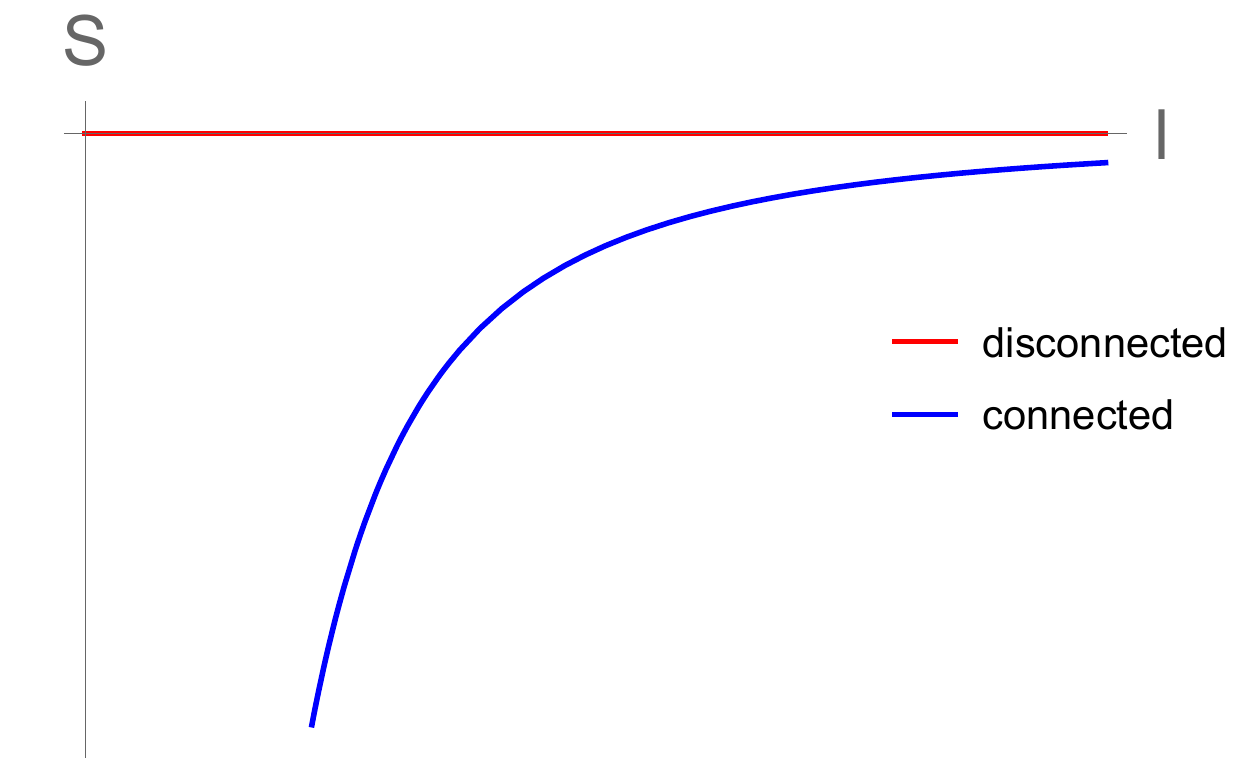}
\caption{\label{fig:1} The EE of the connected and disconnected surfaces}
\end{figure}\\
Figure \ref{fig:1} shows the plot of $S_A$. Since the EE with respect to the disconnected entanglement surface is always larger than the connected one, the EE shows no phase transition.

\section{$AdS_5\times S^5$ with massive flavors}
\label{sec:3}
In this section, we will give a summary of the work \cite{6} (see also in \cite{9}).
Let us consider $AdS_5\times X^5$ first. In the case of $X_5 = S^5$, the Calabi-Yau manifold is 6d Euclidean space and it is dual to $\mathcal{N}=4$ SYM. Under $U(1)$ fibration, the K\"{a}hler-Einstein
base of $S^5$ is $CP^2$, so the metric of this 5d Sasaki-Einstein manifold can be written as:
\begin{equation}
{ds}^2_{X_5}={ds}^2_{KE}+(A_{KE})^2\,,
\end{equation}
where $A_{KE}$  is the connection one form. The metric of $CP^2$ reads
\begin{equation}
{ds}^2_{{CP}^2}=\frac{1}{4}{d\chi}^2+\frac{1}{4}{\cos}^2{\frac{\chi}{2}}({d\theta}^2+{\sin}^2\theta{d\varphi}^2)+\frac{1}{4}{\cos}^2{\frac{\chi}{2}}{\sin}^2{\frac{\chi}{2}}(d\psi+{\cos\theta}d\varphi)^2\,,
\end{equation}
\begin{equation*}
A_{{CP}^2}=\frac{1}{2}{\cos}^2\frac{\chi}{2}(d\psi+{\cos\theta}d\varphi)\,,
\end{equation*}
\begin{equation*}
0\leq\chi,\theta\leq\pi,0\leq\varphi,\tau<2\pi,0\leq\psi<4\pi\,.
\end{equation*}
The D3 branes are put on the tip of a CY cone over $S^5$. The number of D3 branes is $N_c$, thus we get $\mathcal{N}=4$ SYM with gauge group $SU(N_c)$.

Next let us consider the D3/D7 brane system. We introduce D7 branes as matter in our $\mathcal{N}=4$ SYM \cite{5}. The intersection of D3 branes and D7 branes can be seen in the following table~\ref{tab:1}~(see also~\cite{9}):
\begin{table}[hbp]
\centering
\begin{tabular}{|l|l|l|l|l|l|l|l|l|l|l|}
\hline
&t&$x^1$&$x^2$&$x^3$&r&$a^1$&$a^2$&$a^3$&$a^4$&$a^5$\\
\hline
D3&X&X&X&X&&&&&&\\
\hline
D7&X&X&X&X&X&X&X&X&&\\
\hline
\end{tabular}
\caption{\label{tab:1} The intersection of D3 branes and D7 branes}
\end{table}\\
where $a^1\sim a^5$ are the coordinates of the compact manifold $X_5$. D3 branes are located along the radial direction $r$ of AdS space at $r = 0$. The D7 branes which are dual to flavors locate at $r = r_q$ wrapping an $S^3$ inside $S^5$ and spacetime filling in the UV region, where $r_q$ is related to the mass of flavor. When $r_q=0$, that means D3, D7 branes are coincide at the origin, which is related to the massless flavor case. These $N_f$ D7 branes bring a $U(N_f)$ symmetry to the theory. If the number of flavors is not finite, say in the Veneziano limit $N_f, N_c \rightarrow \infty$, where $N_f/N_c$ is fixed, after the smearing procedure the $U(N_f)$ symmetry would become ${U(1)^{N_f}}$ \cite{10}. Notice that the backreacted D7 branes would change the background geometry unlike the case in probe limit. The need for smearing procedure is to avoid the difficulty of the $\delta$ function when computing the integral of DBI action \cite{11}. Notice that now this D3/D7 brane system only preserves $\mathcal{N}=1$ supersymmetry in 4d \footnote{For a detailed review of the addition of unquenched flavor and smearing procedure, one can refer to the paper \cite{12} in which the authors gave a detailed unquenched constructions and studied various models.}.

After applying the smearing procedure, the DBI action without any $\delta$ function reads:
\begin{equation}
S=S_ {\uppercase\expandafter{\romannumeral2}B}+S_{fl}\,,
\end{equation}
where
\begin{equation*}
S_ {\uppercase\expandafter{\romannumeral2}B}=\frac{1}{2\kappa^{2}_{10}}\int{d^{10} x}\sqrt{-g_{10}}[R-\frac{1}{2}\partial_M\Phi\partial^M\Phi-\frac{1}{2}e^{2\Phi}F^{2}_{(1)}-\frac{1}{2}\frac{1}{5!}F^{2}_{(5)}]\,,
\end{equation*}
and the action of D7 branes is
\begin{equation*}
S_{fl}=-T_7\sum_{N_f}\left(\int{d^8x}e^\Phi\sqrt{-g_8}-\int C_8\right)
\end{equation*}
with the gravitational constant of
\begin{equation*}
\frac{1}{\kappa_{10}^{2}}=\frac{T_7}{g_s}=\frac{1}{(2\pi)^2g_{s}^{2}\alpha^{\prime 4}}\,.
\end{equation*}

Now we can write the metric ansatz of the background in zero temperature:
\begin{equation}
{ds}^{2}_{10}=h^{-\frac{1}{2}}(-{dt}^2+{d\vec{x}}^{2}_{3})+h^{\frac{1}{2}}[S^8F^2{d\sigma}^2+S^2{ds}^{2}_{{CP}^2}+F^2(d\tau+A^{2}_{{CP}^2})]\,.
\end{equation}
Notice that the functions $h$, $S$, $F$ appeared in the above metric are only depend on the radial direction coordinate $\sigma$. The next step is to find out the first-order BPS equations to solve all these unknown functions. They are given as follows:
\begin{equation}
\begin{split}
\partial_{\sigma}h=-Q_c \,,
\qquad
\partial_{\sigma}F=S^4F(3-2\frac{F^2}{S^2}-\frac{Q_f}{2}e^{\Phi}\cos^4\frac{\chi_{wv}}{2}) \,,
\\
\partial_{\sigma}S=S^3F^2 \,,
\qquad
\partial_{\sigma}\chi_{wv}=-2S^4\tan\frac{\chi_{wv}}{2}\,,
\qquad
\partial_{\sigma}\Phi=Q_fS^4e^{\Phi}\cos^4\frac{\chi_{wv}}{2}\,.
\end{split}
\end{equation}
Here $Q_c$ and $Q_f$ are proportional to $N_c$ and $N_f$ respectively. Change the coordinate $\sigma$ to $\rho$ by defining $d\rho=S^4d\sigma$.

If $\rho > \rho_q$ where $\rho_q$ is related to the mass of flavor, $S,F,\Phi$ are given by:
\begin{equation}
\begin{split}
S=&\sqrt{{\alpha}^{\prime}}e^{\rho}{(1+\epsilon_\ast(\frac{1}{6}+\rho_\ast-\rho-\frac{1}{6}e^{6\rho_q-6\rho}-\frac{3}{2}e^{2\rho_q-2\rho}+\frac{3}{4}e^{4\rho_q-4\rho}-\frac{1}{4}e^{4\rho_q-4\rho_\ast}+e^{2\rho_q-2\rho_\ast}))}^{\frac{1}{6}}\,,
\\
F=&\sqrt{{\alpha}^{\prime}}e^{\rho}\frac{{(1+\epsilon_\ast(\rho_\ast-\rho-e^{2\rho_q-2\rho}+\frac{1}{4}e^{4\rho_q-4\rho}+e^{2\rho_q-2\rho_\ast}-\frac{1}{4}e^{4\rho_q-4\rho_\ast}))}^{\frac{1}{2}}}{{(1+\epsilon_\ast(\frac{1}{6}+\rho_\ast-\rho-\frac{1}{6}e^{6\rho_q-6\rho}-\frac{3}{2}e^{2\rho_q-2\rho}+\frac{3}{4}e^{4\rho_q-4\rho}-\frac{1}{4}e^{4\rho_q-4\rho_\ast}+e^{2\rho_q-2\rho_\ast}))}^{\frac{1}{3}}}\,,
\\
\Phi=&\Phi_\ast-\log(1+\epsilon_\ast(\rho_\ast-\rho-e^{2\rho_q-2\rho}+\frac{1}{4}e^{4\rho_q-4\rho}+e^{2\rho_q-2\rho_\ast}-\frac{1}{4}e^{4\rho_q-4\rho_\ast}))\,.
\end{split}
\end{equation}
When $\rho_q\rightarrow\infty$, we can recover the massless solution \cite{6}. As for the case of $\rho < \rho_q$, since the flavors do not reach the region of $\rho < \rho_q$, we can get the functions with an additional condition $Q_f = 0$. After applying the continuity condition, they are:
\begin{equation}
S=F=\sqrt{\alpha^\prime}e^{\rho}e^{-\frac{1}{6}(\Phi_{IR}-\Phi_{\ast})}\,,
\end{equation}
\begin{equation}
\Phi_{IR}=\Phi_{q}=\Phi_{\ast}-\log(1+\epsilon_{\ast}(\rho_{\ast}-\rho_q-\frac{3}{4}+e^{2\rho_q-2\rho_{\ast}}-\frac{1}{4}e^{4\rho_q-4\rho_{\ast}}))\,.
\end{equation}
Notice that the dilaton blows up at $\rho = \rho_{LP}$, so the solutions are defined up to $\rho < \rho_{LP}$, where $\rho_{LP}$ is the UV Landau pole. This problem can be solved by sending $\rho_{\ast}\rightarrow\infty$ to decouple the UV Landau pole (see also in \cite{13}). $\Phi_{\ast}$ is the dilaton valued at $\rho_{\ast}$, and $\epsilon_{\ast}=Q_fe^{\Phi_{\ast}}$. Setting the 't Hooft coupling at $\rho_{\ast}$ as $\lambda_{\ast}$,  we have
\begin{equation}
\epsilon_{\ast}=\frac{Vol(X_3)}{16\pi Vol(X_5)}\frac{N_f}{N_c}\lambda_{\ast}\,.
\end{equation}
In our case, when $X_5=S^5$, $\epsilon_{\ast}={\lambda_{\ast}}\frac{1}{8\pi}\frac{N_f}{N_c}$. Using the same method, we can also define
\begin{equation}
\epsilon_{q}=\frac{Vol(X_3)}{16\pi Vol(X_5)}\frac{N_f}{N_c}\lambda_{q}\,,
\end{equation}
which will be used as the expansion parameter. $q$ means that the quantities are calculated at $\rho_q$. From the above expressions, we can find that $\epsilon_q$ and $\epsilon_{\ast}$ are the same in first order of $\epsilon_{\ast}$:
\begin{equation}
\epsilon_q=\epsilon_{\ast}\frac{e^{\Phi_q}}{e^{\Phi_{\ast}}}=\epsilon_{\ast}+O(\epsilon_{\ast}^{2})\,.
\end{equation}

For convenience, we will not solve the differential equation for $h$, but instead we will make an implying coordinate transformation to simplify the expression of $h$:
\begin{equation}
h=\frac{R^4}{r^4},\quad R^4=\frac{1}{4}Q_c=\frac{1}{4}N_c\frac{(2\pi)^4g_s\alpha^{\prime2}}{Vol(X_5)}\,.
\end{equation}
Notice that we are using the same form of the expression of $h$ as in \cite{6}, but different from \cite{7} for the reason that we will compare our result with the result of pure $AdS_5\times S^5$ computed in section \ref{sec:2.3}.
For the case of $\rho > \rho_q$, with the integral constant fixed by $r_\ast=r(\rho_\ast)=\sqrt{\alpha^{'}}e^{\rho_\ast}(1+\frac{1}{720}\epsilon_q(15e^{4\rho_q-4\rho_\ast}-8{e^{6\rho_q-6\rho_\ast}}))$, $r(\rho)$ reads:
\begin{equation}
\begin{split}
r(\rho)=&\sqrt{\alpha^{\prime}}e^\rho(1+\epsilon_q(\frac{1}{6}\rho_{\ast}-\frac{1}{6}\rho-\frac{1}{72}-\frac{1}{90}{e^{6\rho_q-6\rho}}+\frac{1}{16}{e^{4\rho_q-4\rho}}-\frac{1}{24}{e^{4\rho_q-4\rho_\ast}}-\frac{1}{6}{e^{2\rho_q-2\rho}}\\
+&\frac{1}{6}{e^{2\rho_q-2\rho_\ast}}+\frac{1}{72}e^{4\rho-4\rho_\ast}))\,,
\end{split}
\end{equation}
and the functions of $S(r)$, $F(r)$ and $\Phi(r)$ are:
\begin{equation}
S(r)=r(1+\frac{\epsilon_q}{720}(-12\frac{r^{6}_{q}}{r^6}+45\frac{r^{4}_{q}}{r^4}-60\frac{r^{2}_{q}}{r^2}+30-10\frac{r^4}{r^{4}_{\ast}}))\,,
\end{equation}
\begin{equation}
F(r)=r(1+\frac{\epsilon_q}{720}(48\frac{r^{6}_{q}}{r^6}-135\frac{r^{4}_{q}}{r^4}+120\frac{r^{2}_{q}}{r^2}-30-10\frac{r^4}{r^{4}_{\ast}}))\,,
\end{equation}
\begin{equation}
\label{eq:3.16}
\Phi(r)=\Phi_\ast+\epsilon_q(\log\frac{r}{r_\ast}+\frac{r^{2}_{q}}{r^2}-\frac{1}{4}\frac{r^{4}_{q}}{r^4}-\frac{r^{2}_{q}}{r^{2}_{\ast}}+\frac{1}{4}\frac{r^{4}_{q}}{r^{4}_{\ast}})\,.
\end{equation}
Notice that in (\ref{eq:3.16}), when $r_q=0$ we recover the flavorless case in \cite{6}.
For the case of $\rho < \rho_q$, after applying the continuity condition at $r=r(\rho_q)$, $r(\rho)$ reads:
\begin{equation}
r(\rho)=\sqrt{\alpha^{\prime}}e^\rho(1+\epsilon_q(\frac{\rho_{\ast}}{6}-\frac{\rho_q}{6}-\frac{1}{24}{e^{4\rho_q-4\rho_{\ast}}}+\frac{1}{6}e^{2\rho_q-2\rho_\ast}-\frac{1}{8}-\frac{1}{240}e^{4\rho-4\rho_q}+\frac{1}{72}e^{4\rho-4\rho_\ast}))\,,
\end{equation}
and the functions of $S(r)$, $F(r)$ and $\Phi(r)$ are:
\begin{equation}
S(r)=F(r)=r(1+\epsilon_q\frac{1}{720}(3\frac{r^4}{r^{4}_{q}}-10\frac{r^4}{r^{4}_{\ast}}))\,,
\end{equation}
\begin{equation}
\Phi(r)=\Phi_\ast+\epsilon_q(\log\frac{r_q}{r_\ast}-\frac{3}{4}-e^{2\rho_q-2\rho_\ast}+\frac{1}{4}e^{4\rho_q-4\rho_\ast})\,.
\end{equation}
The above solutions also depend on a scale $r_\ast$ which is related to the UV Landau pole. As explained in \cite{6}, the perturbative results are only available when $r<r_\ast$, hence we need to send $r_\ast\rightarrow+\infty$ at the end of the computation.

The metric of $AdS_5\times S^5$ with massive flavors in Veneziano limit with radial coordinate $r$ is:
\begin{equation}
\label{eq:3.17}
{ds}^{2}_{10}=h^{-\frac{1}{2}}(-{dt}^2+{d\vec{x}}^{2}_{3})+h^{\frac{1}{2}}[F^2\frac{S^8}{r^{10}}{dr}^2+S^2{ds}^{2}_{{CP}^2}+F^2(d\tau+A^{2}_{{CP}^2})]\,.
\end{equation}

\section{Computation of entanglement entropy}
\label{sec:4}
We will compute the entanglement length and EE of the flavored $AdS_5\times S^5$ with massive flavors with the shape of a strip in this section. From~(\ref{eq:3.17}), we get:
\begin{equation}
\beta(r)=hF^2S^8\frac{1}{r^{10}},\quad H(r)=hF^2S^8\,.
\end{equation}
$\tilde{r}$ is the turning point along the radial direction, which has the same definition as in section~\ref{sec:2.1}. All the functions with tilde defined below mean that they are valued at $\tilde{r}$.

Comparing to~(\ref{eq:2.15}), the entanglement length is:
\begin{equation}
l=2\int^{+\infty}_{\tilde{r}}dr\frac{\sqrt{h\tilde{h}}F{\tilde{F}}S^4{\tilde{S}}^4}{r^5\sqrt{hF^2S^8-\tilde{h}{\tilde{F}}^2{\tilde{S}}^8}}\,.
\end{equation}
The area of the connected minimal surface is then:
\begin{equation}
Area_C=2L^2\int^{+\infty}_{\tilde{r}}dr\frac{h^{\frac{3}{2}}F^3S^{12}}{r^5\sqrt{hF^2S^8-\tilde{h}{\tilde{F}}^2{\tilde{S}}^8}}\,,
\end{equation}
while the disconnected one is:
\begin{equation}
Area_D=2L^2\int^{+\infty}_{0}dr\frac{hF^2S^{8}}{r^5}\,.
\end{equation}
We have divided the result of area by $Vol(S^5)$. As mentioned above, we will simplify the computation by regarding the warp factor $h$ as~$\frac{R^4}{r^4}$. To find the entanglement length, we will expand our result up to first order of the expansion parameter $\epsilon_q$.

Comparing to section~\ref{sec:2.3}, the computation gets much complex after the introduction of backreacted massive flavors. Notice that the turning point of the entanglement surface $\tilde{r}$ may be larger or smaller than $r_q$ which is related to the mass of the flavors. When~$\tilde{r}<r_q$ which means that the entanglement surface extends to the transverse space where the D7 branes vanish, we need to do the integration from $\infty$ to $r_q$, then from $r_q$ to $\tilde{r}$.

Before the explicit computation, let us quickly check the behavior of the entanglement length $l$ by using the function $\mathcal{Y}(\tilde{r})$ defined in (3.1) of \cite{15}:
\begin{equation*}
\mathcal{Y}(\tilde{r})=2\pi\frac{H(r)\sqrt{\beta(r)}}{H'(r)}\bigg|_{r=\tilde{r}}\,.
\end{equation*}
\begin{equation}
\begin{split}
\label{eq:4.5}
l(\tilde{r}\rightarrow+\infty)\sim\mathcal{Y}(\tilde{r}\rightarrow+\infty)\sim \lim_{\tilde{r} \to +\infty}\frac{(\tilde{r}^6+\tilde{r}^2)\sqrt{\frac{1}{\tilde{r}^4}+\frac{1}{\tilde{r}^8}}}{\tilde{r}^5+\tilde{r}}\sim \lim_{\tilde{r} \to +\infty}\frac{1}{\tilde{r}}\,,
\\
l(\tilde{r}\rightarrow0)\sim \mathcal{Y}(\tilde{r}\rightarrow0)\sim\lim_{\tilde{r} \to 0}\frac{\tilde{r}^6}{\tilde{r}^5}\sqrt{\frac{1}{\tilde{r}^4}}\sim\lim_{\tilde{r} \to 0}\frac{1}{\tilde{r}}\,.
\end{split}
\end{equation}

For the case when $\tilde{r}>r_q$, the entanglement length is given by:
\begin{equation}
\label{eq:4.6}
\begin{split}
{l(\tilde{r})}_>=&2{\int}^{+\infty}_{\tilde{r}}dr\frac{R^2{\tilde{r}}^3}{r^2\sqrt{r^6-\tilde{r}^6}}(1+\epsilon_q(\frac{r^{4}_{q}}{16}(\frac{1}{r^2}+\frac{1}{\tilde{r}^2})-\frac{r^{2}_{q}}{6}(\frac{1}{r^2}+\frac{1}{\tilde{r}^2})+\frac{1}{4}-\frac{5}{36}\frac{r^4}{r^{4}_{\ast}}\\
&-\frac{1}{r^6-\tilde{r}^6}(\frac{r^{4}_{q}}{16}(r^2-\tilde{r}^2)-\frac{r^{6}_{q}}{6}(r^4-\tilde{r}^4)+\frac{1}{8}(r^6-\tilde{r}^6)-\frac{5}{144r^{4}_{\ast}}(r^{10}-r^{10}_{\ast}))))\\
=&\frac{2R^2}{\tilde{r}}(\frac{\sqrt{\pi}\Gamma(\frac{2}{3})}{\Gamma(\frac{1}{6})}+\epsilon_q(\frac{1}{8}\frac{\sqrt{\pi}\Gamma(\frac{2}{3})}{\Gamma(\frac{1}{6})}+\frac{r^{4}_{q}}{{\tilde{r}}^{4}}(\frac{1}{90}\frac{\sqrt{\pi}\Gamma(\frac{1}{3})}{\Gamma(\frac{5}{6})}-\frac{1}{3}\frac{\sqrt{\pi}\Gamma(\frac{2}{3})}{\Gamma(\frac{1}{6})})\\
&-\frac{r^{2}_{q}}{{\tilde{r}}^2}(\frac{1}{9}-\frac{1}{18}\frac{\sqrt{\pi}\Gamma(\frac{2}{3})}{\Gamma(\frac{1}{6})})))+2{\int}^{+\infty}_{\tilde{r}}dr\frac{R^2{\tilde{r}}^3}{r^2\sqrt{r^6-\tilde{r}^6}}(-\frac{5}{36}\frac{r^4}{r^{4}_{\ast}}+\frac{5}{144}\frac{r^{10}-\tilde{r}^{10}}{r^{4}_{\ast}(r^6-\tilde{r}^6)})\epsilon_q\,.
\end{split}
\end{equation}
Notice that the last integral term in (\ref{eq:4.6}) includes an extra scale $r_\ast\rightarrow+\infty$, $r<r_\ast$. This term appears to be the correction to the term of $2{\int}^{+\infty}_{\tilde{r}}dr\frac{R^2{\tilde{r}}^3}{r^2\sqrt{r^6-\tilde{r}^6}}\frac{1}{4}\epsilon_q$ which is of order $1/\tilde{r}$. When $\tilde{r}\rightarrow+\infty$, $l(\tilde{r})$ goes to zero.

In the expression of the entanglement length above, we have two constants: one is the expansion parameter $\epsilon_q$ which is of order 1, and the other is $r_q$ which is related to the mass of the flavors.
Both of them can be set to 1 since we do not consider the influence of the quark mass and the ratio of flavor and color branes in this paper.
For the reason that the EE is proportional to the area of the entanglement surface up to a constant by the area law, we will study the property of the area instead of the exact value of the EE.

The area of the connected minimal surface when~$\tilde{r}>r_q$ is:
\begin{equation}
\begin{split}
(Area_C)_>&=2L^2R^4\int^{+\infty}_{1}dx\frac{x^4{\tilde{r}}^2}{\sqrt{x^6-1}}(1+\epsilon_q(\frac{3}{16}\frac{r^{4}_{q}}{{\tilde{r}}^4}\frac{1}{x^4}-\frac{1}{2}\frac{r^{2}_{q}}{{\tilde{r}}^2}\frac{1}{x^4}+\frac{1}{4}\\
&-\frac{5}{36}\frac{r^4}{r^{4}_{\ast}}-\frac{1}{x^6-1}(\frac{1}{16}\frac{r^{4}_{q}}{{\tilde{r}}^4}(x^2-1)-\frac{1}{6}\frac{r^{2}_{q}}{{\tilde{r}}^2}(x^4-1))))\,.
\end{split}
\end{equation}
Notice that the area is divergent in the UV region.
The area of the disconnected minimal surface is:
\begin{equation}
\label{eq:4.8}
\begin{split}
Area_D&=2L^2R^4\int^{+\infty}_{0}{dr}r(1+\epsilon_q(\frac{1}{8}\frac{r^{4}_{q}}{r^4}-\frac{1}{3}\frac{r^{2}_{q}}{r^2}+\frac{1}{4}-\frac{5}{36}\frac{r^4}{r^{4}_{\ast}}))\\
&=\left.L^2R^4(r^2+\epsilon_q(r^2(\frac{1}{4}-\frac{5}{216}\frac{r^4}{r^{4}_{\ast}})-\frac{2}{3}r^{2}_{q}\log r-\frac{1}{8}\frac{r^{2}_{q}}{r^2}))\right|^{\infty}_{0}\,.
\end{split}
\end{equation}
Notice that we have another divergent part in~(\ref{eq:4.8}) with leading order divergence of $\frac{1}{r^2}$ in IR region~($r\rightarrow0$).
The counterterm we used for canceling the divergent part~($a\rightarrow+\infty$) is:
\begin{equation}
\label{eq:4.9}
L^2R^4(a^2+\epsilon_q(\frac{49}{216}a^2-\frac{2}{3}r^{2}_{q}\log a))\,.
\end{equation}
Following the same procedure in section~\ref{sec:2.3}, we can get a UV cut-off independent area which is given by:
\begin{equation}
\begin{split}
Area_>&=L^2R^4{\tilde{r}}^{2}(-\frac{\sqrt{\pi}\Gamma(\frac{2}{3})}{\Gamma(\frac{1}{6})}+\epsilon_q(-\frac{1}{4}\frac{\sqrt{\pi}\Gamma(\frac{2}{3})}{\Gamma(\frac{1}{6})}+\frac{1}{144}\frac{r^{4}_{q}}{{\tilde{r}}^4}(\frac{8\sqrt{\pi}\Gamma(\frac{1}{3})}{\Gamma(\frac{5}{6})}-\frac{6\sqrt{\pi}\Gamma(\frac{2}{3})}{\Gamma(\frac{1}{6})})\\
&+\frac{1}{144}\frac{r^{2}_{q}}{{\tilde{r}}^2}(\frac{16\sqrt{\pi}\Gamma(\frac{2}{3})}{\Gamma(\frac{1}{6})}+16(6\log\tilde{r}-1-2\log 2))))\,.
\end{split}
\end{equation}
Notice that if $\epsilon_q=0$, which is the unflavored case, we can recover the result of~(\ref{eq:2.24}). When $\tilde{r}\rightarrow+\infty$, the leading-order divergence of the area is:~$Area_>\sim-{\tilde{r}}^2$. This behavior coincides with that of the unflavored one. Notice that $Area_D$ is divergent even after subtracting the counterterm, thus $Area_C$ is always smaller than $Area_D$ when $\tilde{r}>r_q$.

If~$\tilde{r}<r_q$, the minimal surface extends into the transverse space much further, but the behavior in UV region does not change, so the counterterm remains the same.
For the case when~$\tilde{r}<r_q$, we will compute them numerically. The divergent part of the area in the UV region is the same as (\ref{eq:4.9}). We will set the quantity related to the mass of flavors $r_q$ equal to 1. In the IR region, the terms with the form of $r^4/r^{4}_{\ast}$ in the functions of $F(r)$ and $S(r)$ vanish since $r\ll r_\ast\rightarrow+\infty$.
\begin{figure}[bhtp]
\centering
\includegraphics[width=.45\textwidth,clip]{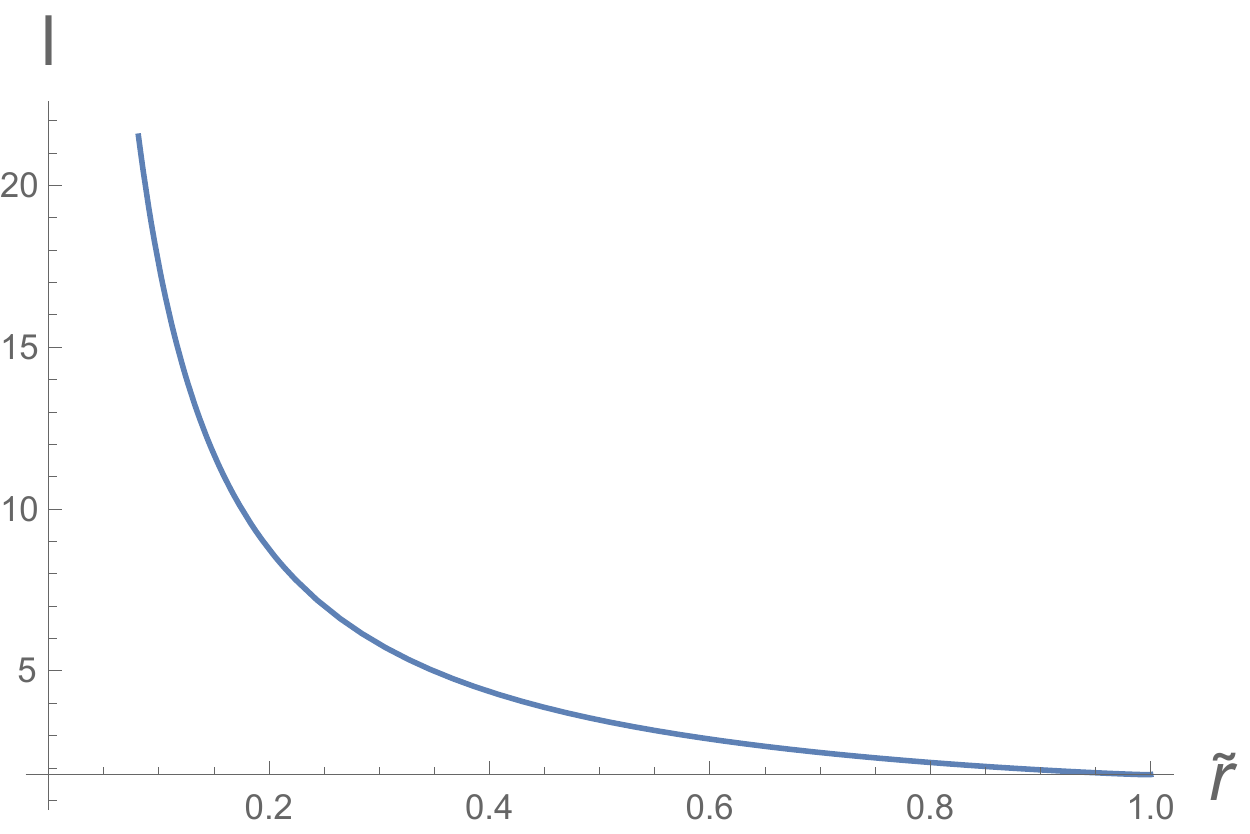}
\caption{\label{fig:2} The entanglement length of the connected minimal surface when~$\tilde{r}<r_q$}
\end{figure}\\
\begin{figure}[bhtp]
\centering
\includegraphics[width=.45\textwidth,clip]{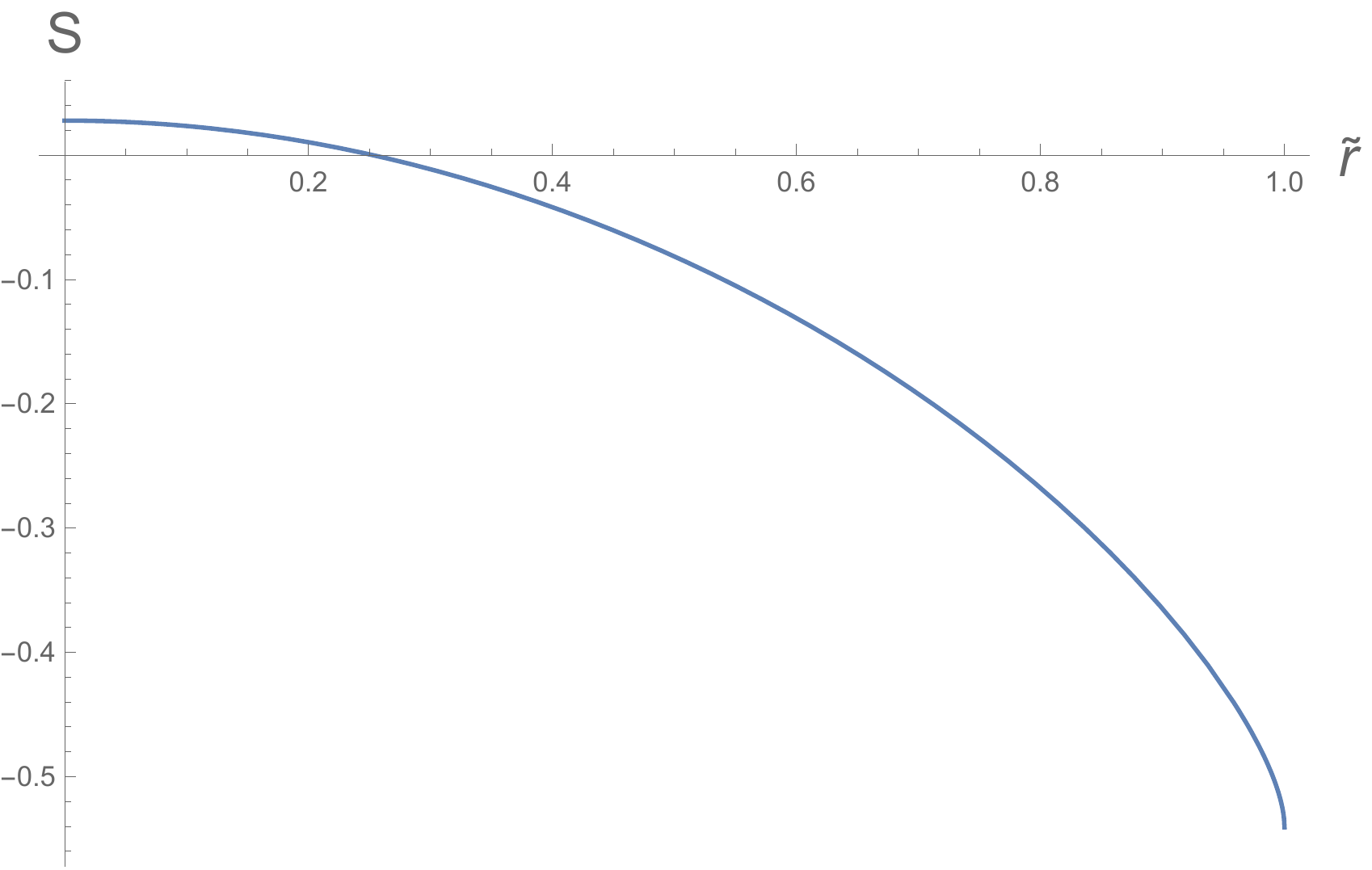}
\caption{\label{fig:3} The EE of the connected minimal surface when~$\tilde{r}<r_q$}
\end{figure}\\
The leading order of the length is:~$l\sim 1/{\tilde{r}}$ (the behavior of $l$ is the same with the behavior of the approximation function $\mathcal{Y}$ analyzed in (\ref{eq:4.5})), while that of the area is:~$Area_<\sim-{\tilde{r}}^2$. These can be seen also in figure~\ref{fig:2}~and~\ref{fig:3}.~(As mentioned above, the constants~$\epsilon_q$~and~$r_q$~are set to~1, and $R$ is also been ignored.) Comparing with the discussion in section~\ref{sec:2.3}, notice that even after the introduction of the massive flavor, the behavior of leading order when $\tilde{r}\rightarrow0$ remains the same except that there is a correction at the origin due to the addition of the flavors. Notice also that $Area_C$ is always smaller than $Area_D$ when $\tilde{r}<r_q$ for the same reason when $\tilde{r}>r_q$.

From all the discussion in this section, we can safely come to a conclusion that: in the backreacted $AdS_5\times S^5$ with massive flavors, the entanglement length and EE have the same behavior as the unflavored $AdS_5\times S^5$ when $\tilde{r}\rightarrow0$ and $\tilde{r}\rightarrow+\infty$. The length $l(\tilde{r})$ ranges from 0 to infinity, thus there is no maximum of $l$. When $\tilde{r}\rightarrow0$, $l$ diverges; when $\tilde{r}\rightarrow+\infty$, $l\rightarrow0$.  When $\tilde{r}\rightarrow+\infty$, $Area_C(\tilde{r})\rightarrow-\infty$; when $\tilde{r}\rightarrow0$, $Area_C(\tilde{r})$~is closed to the origin. The EE which is proportional to the area $Area_C(\tilde{r})$ shares the same behavior with the area.

In~\cite{1}, the EE as function of $l$ has a phase transition behavior which is called "butterfly-shape" due to the double-valueness of $S(l)$. The phase transition occurs since the difference of connected and disconnected minimal surfaces changes sign in all the cases the authors studied. In the backreacted $AdS_5\times S^5$ with massive flavors which is the case we study, considering the EE as a function of entanglement length $l$, from the discussion above, there exists no double-valueness in $Area_C(l)$. Since $l$ ranges from 0 to $\infty$, when $l\rightarrow+\infty$ which only corresponds to $\tilde{r}\rightarrow0$, we can get that $Area_C(l)$ is close to zero; when $l\rightarrow0$ which is only correspondent to $\tilde{r}\rightarrow+\infty$, we get that $Area_C(l)\rightarrow-\infty$. The behavior of the function $Area_C(l)$ is of order~$Area_C(l)\sim -1/{l}^2$. This result is similar to that in figure~\ref{fig:1}. Comparing with (\ref{eq:4.8}), we find that the area of disconnected minimal surface is always larger than that of the connected one.  I am unable to find the exact function $Area_C(l)$, but the above discussion is enough for us to conclude that there is no phase transition in the backreacted $AdS_5\times S^5$ with massive flavors and the EE related to the connected minimal surface always remains the dominant one.

\section{Conclusion}
\label{sec:5}
In this paper we studied the entanglement entropy of $AdS_5 \times S^5$ with massive flavors in the Veneziano limit\footnote{Recent study of entanglement entropy to probe the confinement/deconfinement phase transitions in holographic models can be found in \cite{17}.}. Dividing one of the spatial direction into a line segment with length $l$, we get two complementary subsystems: A and its complement B. We calculate the EE between them by using the method of~\cite{1}. After removing the UV divergent term of the EE, we get universal UV cut-off independent term. (For other choices of the universal term, see~\cite{16}.)

In section~\ref{sec:2}, we calculate the entanglement length and EE of pure $AdS_5\times S^5$, finding that the entanglement length is divergent at the origin and the EE of the disconnected minimal surface is always larger than the connected one, thus the lower EE related to the connected minimal surface is always preferred, hence there is no phase transition between them. In section~\ref{sec:4}, we calculate the entanglement length and EE of the $AdS_5\times S^5$ with $N_f$ massive flavors at zero temperature. We find that there is no phase transition even after introducing the D7 branes as matter into our theory. The reason of it maybe lies in that we are considering the model in zero temperature. We will study the non-zero temperature case in the future and see if we can get something different.

Let us close this section by discussing some possible future work. It would be interesting to compare the confinment/deconfinement phase transition phenomenon in flavored~$AdS_5\times S^5$ with that of the Klebanov-Witten, Klebanov-Tseytlin, Klebanov-Strassler geometries in the meaning of their dual field theories. We also notice that the procedure during computing the EE is similar to that of the Wilson loop. For review of the Wilson loop of the unquenched dynamical flavor models, one can refer to \cite{18,19} in which the authors developed methods analogous to \cite{1} to study the phase transition phenomenon. The action using for computing the EE is the area action, while that of the Wilson loop is the Nambu-Goto action. The physical meaning of the connection between them still remains unknown (for the numerical study of this problem, one can refer to \cite{15}). Maybe one can try to find the underlying relationship between them.
\acknowledgments

We would like to thank Niko Jokela and Subhash Mahapatra for communications about this paper and especially Carlos Nunez for his useful comment on this paper. G.W. would like to thank his supervisor Prof. Sen Hu for useful discussions and guidance. This work is supported by the Wu Wen-Tsun Key Laboratory of Mathematics at USTC of Chinese Academy of Sciences. The work is partially supported by NSFC under the contract nunber of 11571336.

\end{document}